%% file: main.tex
\documentclass[review]{elsarticle}

\journal{Vehicular Communications}
\usepackage{booktabs}
\usepackage{amstext}
\usepackage{amssymb}
\usepackage{amsmath}
\usepackage{hyperref}

\usepackage{shortcuts}

\begin{document}

\begin{frontmatter}
    \title{\thetitle}
    
    \author[address_ku]{Seonghoon Jeong}
    \ead{seonghoon@korea.ac.kr}
    \author[address_etri]{Boosun Jeon}
    \ead{bsjeon@etri.re.kr}
    \author[address_etri]{Boheung Chung}
    \ead{bhjung@etri.re.kr}
    \author[address_ku]{Huy Kang Kim\corref{mycorrespondingauthor}}
    \cortext[mycorrespondingauthor]{Corresponding author}
    \ead{cenda@korea.ac.kr}
    \address[address_ku]{School of Cybersecurity, Korea University, Seoul 02841, Republic of Korea}
    \address[address_etri]{Cyber Security Research Division, Electronics and Telecommunications Research Institute, Daejeon 34129, Republic of Korea}
    
    \begin{abstract}
        \input{paper/00_abs.tex}
    \end{abstract}
    
\end{frontmatter}


\input{paper/01_introduction.tex}
\input{paper/02_preliminaries.tex}
\input{paper/03_threats.tex}
\input{paper/04_system_design}
\input{paper/05_experiment_result}
\input{paper/06_discussion}

\input{paper/07_related_work}
\input{paper/08_conclusion}

\section*{Acknowledgements}
This work was supported by Institute for Information \& communications Technology Promotion(IITP) grant funded by the Korea government(MSIT) (No.2018-0-00312, Developing technologies to predict, detect, respond, and automatically diagnose security threats to automotive Ethernet-based vehicle).


\input{main.bbl}
\end{document}

%% file: paper/00_abs.tex
Connected and autonomous vehicles (CAVs) are an innovative form of traditional vehicles. Automotive Ethernet replaces the controller area network and FlexRay to support the large throughput required by high-definition applications. As CAVs have numerous functions, they exhibit a large attack surface and an increased vulnerability to attacks. However, no previous studies have focused on intrusion detection in automotive Ethernet-based networks. In this paper, we present an intrusion detection method for detecting audio-video transport protocol (AVTP) stream injection attacks in automotive Ethernet-based networks. To the best of our knowledge, this is the first such method developed for automotive Ethernet. The proposed intrusion detection model is based on feature generation and a convolutional neural network (CNN). To evaluate our intrusion detection system, we built a physical BroadR-Reach-based testbed and captured real AVTP packets. The experimental results show that the model exhibits outstanding performance: the F1-score and recall are greater than 0.9704 and 0.9949, respectively. In terms of the inference time per input and the generation intervals of AVTP traffic, our CNN model can readily be employed for real-time detection.

%% file: paper/01_introduction.tex
\section{Introduction}
Connected and autonomous vehicles (CAVs) are an innovative form of \textit{traditional vehicle}. A significant difference between traditional vehicles and CAVs is the following: In CAVs, a physical (PHY) layer is used for the in-vehicle network (IVN), and automotive Ethernet replaces the controller area network (CAN) and FlexRay. As a result, CAVs support the large communication throughput required by high-definition applications such as video-on-demand, intelligent transport systems, and advanced driver assistance systems (ADAS), which consume various sensor data and video streams \cite{Hank2013AutomotiveMobility}. The IEEE 1722 audio-video transport protocol (AVTP) is the essential mechanism by which automotive Ethernet ensures the reliable transport of time-sensitive and prioritized traffic (e.g., audio and video streaming, as the name suggests). Moreover, the IEEE 1722-2016 standard defines the AVTP for the transmission of time-sensitive control streams (including CAN and FlexRay messages) via automotive Ethernet; the latter was defined by the IEEE P1722a working group \cite{ieee1722-2016}. Accordingly, we consider that AVTP will be one of the most crucial protocols for IVNs in CAVs.%

Currently, cyber-attacks against vehicles are rare (even with a lack of protection mechanisms) because the vehicles have inflexible applications and limited processing resources that are only sufficient for their mobility \cite{Loukas2019}. However, the attack surface and feasibility of attacks are substantial for CAVs because of their additional functions. For example, Automotive Grade Linux or Android OS on an infotainment device could inherit vulnerabilities or malware inbound owing to the connectivity of CAVs \cite{Panarotto2018}. Intrusion detection is a useful method to prepare for cyber-physical attacks on IVNs. Although there have been several studies on intrusion detection systems (IDSs) for IVN security  \cite{Loukas2019, Wu2019}, intrusion detection for systems employing automotive Ethernet has not been studied. The interruption of a media stream owing to intrusion raises not only a usability issue but also security and safety issues in CAVs. Therefore, it is necessary to develop an IDS to address the gap in intrusion detection research with regard to automotive Ethernet.%

Convolutional neural networks (CNNs) are artificial neural networks that are often used in network traffic classification problems (e.g., classifying application protocols or intrusion detection). CNNs are extensively used because they can easily learn and predict one- or multidimensional data using a few parameters. In a CNN, a few sets of convolutional layers and pooling layers extract features from a given input. Finally, dense layers act as an estimator that returns a predicted class or a value using the extracted features. Thus, CNNs simplify supervised learning without the need for much preprocessing of the input data.%

In this paper, we propose an intrusion detection method using a deep learning model. Our method includes the feature generation process and a two-dimensional CNN (2DCNN) model. We designed the feature generator considering the observed characteristics of real AVTP traffic. The detection model distinguishes whether AVTP packets transmitted over automotive Ethernet are benign or injected on a packet-by-packet basis. Finally, we implemented an IDS using the proposed method to evaluate the performance of our method.%

The main contributions of this study are as follows:%

\begin{itemize}
    \item We introduce six possible attack scenarios for automotive Ethernet. We also demonstrate an AVTP replay attack using real audio-video bridging (AVB) devices.
    \item We propose a novel intrusion detection method designed for automotive Ethernet. Our approach comprises the feature generator and the 2D-CNN model. To the best of our knowledge, this is the first attempt to solve the problem of the security of automotive Ethernet-based IVNs.
    \item We implement an IDS to evaluate the performance of our intrusion detection method by using a real dataset that we captured from a BroadR-Reach network. The evaluation results show that the IDS correctly classifies almost all AVTP packets, with very few false negatives. Furthermore, we confirm that our IDS is suitable for real-time detection.
    \item We provide automotive Ethernet intrusion datasets used for our experiment in the PCAP format. In the datasets, each AVTP packet is labeled as ``benign'' or ``injected.'' Readers can access our datasets in \cite{dataset} to reproduce our experiment results, as well as use the datasets in further studies.%

\end{itemize}

The paper is organized as follows. \autoref{sec:preliminaries} introduces the background of automotive Ethernet and AVTP. \autoref{sec:threats} provides plausible attack scenarios targeted to automotive Ethernet. \autoref{sec:system_design} describes the results of our replay attack performed on our real automotive Ethernet network as well as the proposed IDS design---including the feature generator and the artificial neural network used to classify AVTP packets. \autoref{sec:experiment} shows the experiment results on the real AVTP traffic. \autoref{sec:discussion} discusses the AVTP intrusion dataset, some limitations, and remediation strategies. \autoref{sec:related_work} reviews previous studies on IDS designed for CAVs. Finally, \autoref{sec:conclusion} concludes the paper. %

%% file: paper/02_preliminaries.tex
\section{Preliminaries}
\label{sec:preliminaries}
\subsection{Background of emerging automotive Ethernet}
The CAN protocol is widely adopted for IVNs of traditional vehicles: its arbitration mechanism enables a simple bus topology, and its real-time message prioritization considers the importance of each application. However, there are two drawbacks to CAN-based IVNs: (1) security issues and (2) limited bandwidth which is not suitable for various multimedia streams (such as camera data and high-quality audio/video).
\textit{Traditional Ethernet} was once considered a successor of the CAN bus for IVNs. Ethernet is commonly used in local-area networks and provides efficient, low-cost, and high-bandwidth communication. However, 
the best-effort transmission that Ethernet provides is insufficient for CAVs: an unexpected delay or further traffic loss can be highly detrimental in vehicular applications. Thus, Ethernet protocols require some modifications in the PHY and the higher layers to enable their application to IVNs. Consequently, the concept of automotive Ethernet has emerged, and automotive Ethernet has been developed by standardization groups and industry experts.%

\subsection{Automotive Ethernet and IEEE 1722}
\begin{figure}[t]
    \centering
    \includegraphics{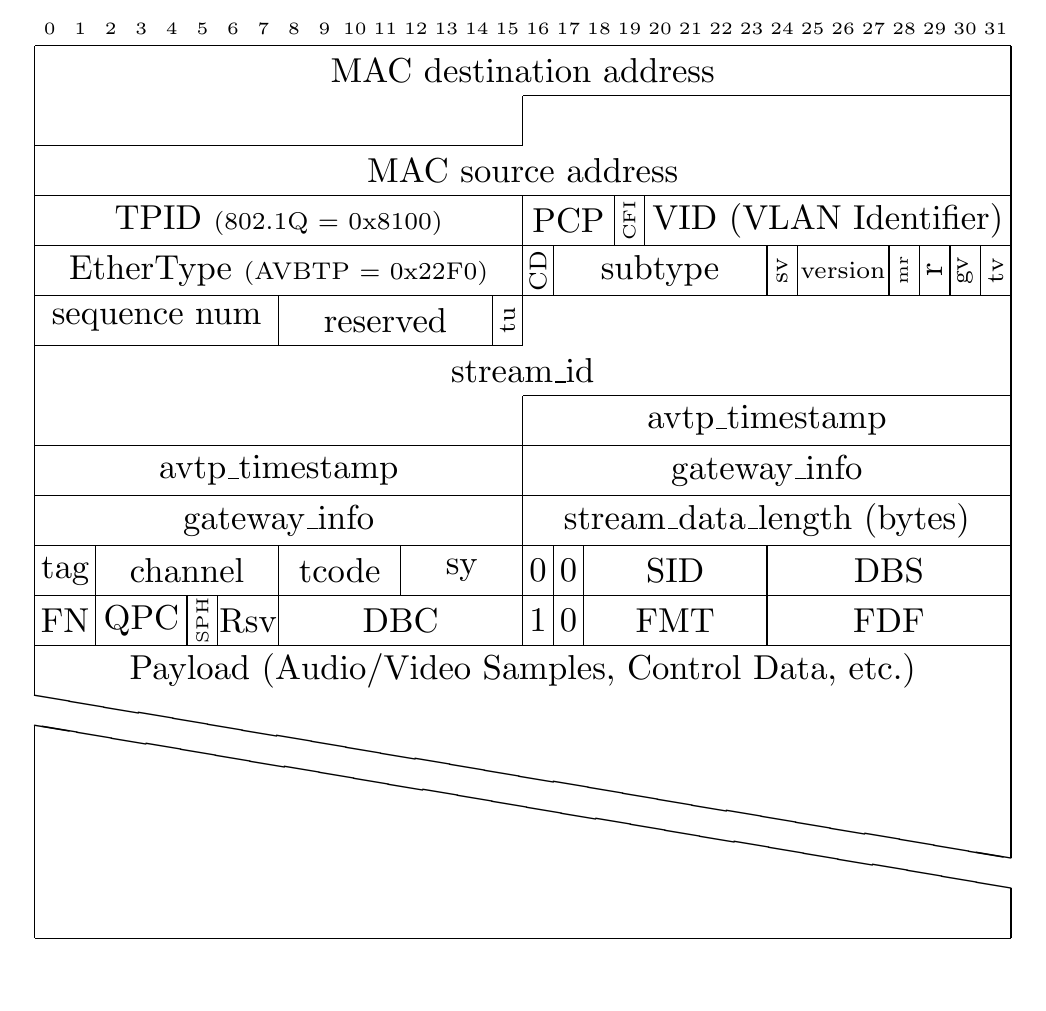}
    \caption{Structure of stream AVTPDU  --- Values of \texttt{TPID}, \texttt{EtherType}, and \texttt{CD} fields are set to 8100h, 22F0h, and 0b, respectively.}
    \label{fig:header}
\end{figure}

Automotive Ethernet has innovations in the PHY and higher layers. Open Alliance's special interest group designed and published BroadR-Reach technology by modifying the PHY employed in Ethernet to meet electrical requirements for automotive Ethernet. Subsequently, IEEE standardized 100BASE-T1, which succeeded BroadR-Reach with a few minor modifications. As a result, automotive Ethernet-based IVNs can be deployed with a twisted-pair cable; furthermore, automotive Ethernet exhibits a high bandwidth of up to 100 Mbps. However, changes in the PHY do not guarantee low-latency, time-sensitive, and prioritized communication based on existing protocols.%

To guarantee the aforementioned features, Avnu Alliance implemented a set of technical standards named AVB. Notably, AVB enables the reliable transport of time-critical streams on automotive Ethernet. Specifically, four protocols---stream reservation protocol, generalized precision time protocol (gPTP), forwarding and queuing for time-sensitive streams, and AVTP---comprise AVB technology. The protocol headers are encapsulated in standard Ethernet headers when the packets are transmitted through automotive Ethernet. For example, an AVTP packet is encapsulated by the general MAC header as well as the VLAN header. The VLAN header functions as the prioritization enabler for automotive Ethernet; in addition, it enables network segmentation, as intended. With the support of these protocols, AVB-compatible devices become synchronized with each other and produce or consume streams regardless of topological complexity \cite{autoethdefinitiveguide}.%

\autoref{fig:header} shows the structure of a stream AVTP data unit (AVTPDU). AVTP is in charge of delivering the application payload. The payload can include audio and video streams, in addition to CAN or FlexRay messages for controlling in-vehicle electronic control units (ECUs). AVTPDU refers to an AVTP packet; AVTPDUs are categorized into \textit{control AVTPDU} (value of the \texttt{CD} field is set as 1) and \textit{stream AVTPDU} (value of the \texttt{CD} field is set as 0). AVB devices send control AVTPDUs to discover themselves or establish/disconnect a stream session. Meanwhile, in-vehicle switches learn the positions of AVTP listeners. After the session is established, the AVB talker transmits stream AVTPDUs containing application data. Then, the in-vehicle switches multicast the stream AVTPDUs. Stream AVTPDUs always adhere to a VLAN tag to represent their priority among less time-critical packets. The \texttt{channel} field specifies the type of protocol for the subsequent \texttt{Payload} field. %

%% file: paper/03_threats.tex
\section{Threats in automotive Ethernet}
\label{sec:threats}
\begin{figure}[t]
    \centering
    \includegraphics[width=\linewidth]{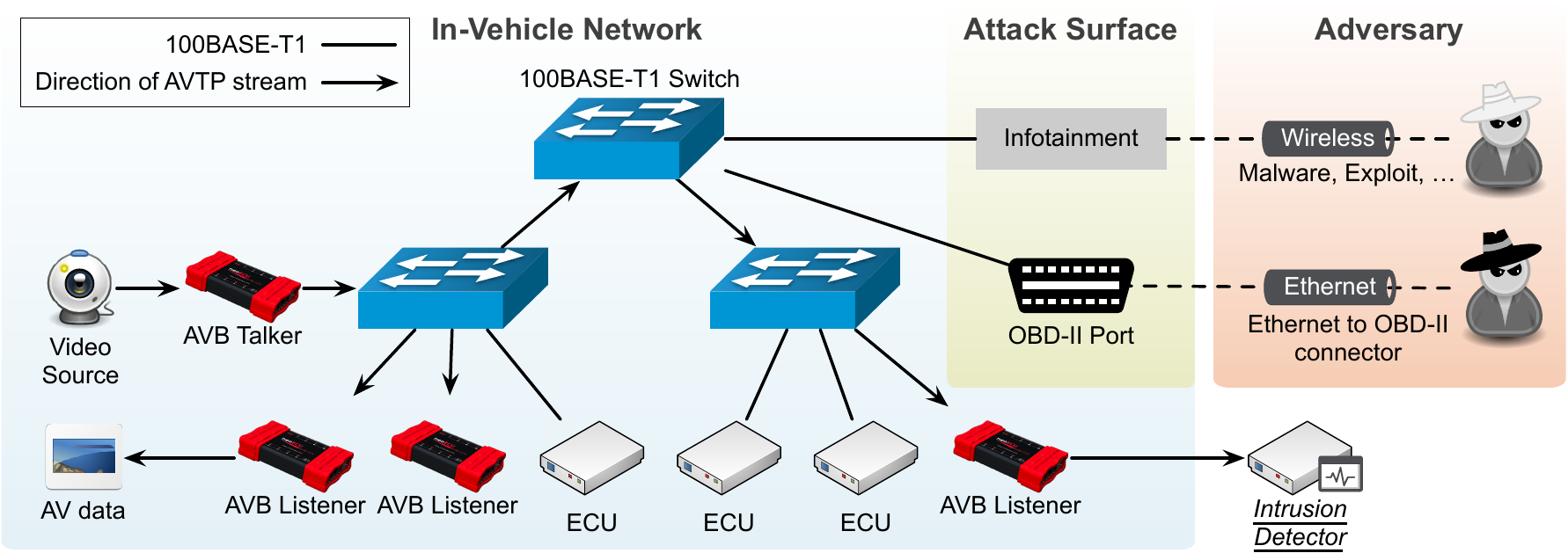}
    \caption{Example of in-vehicle automotive Ethernet topology along with possible packet injection attack scenarios. AVTP streams flow from the AVB talker to the AVB listeners.}
    \label{fig:topology}
\end{figure}

Several studies have evaluated threats to IVNs, especially with regard to the CAN bus. In contrast, there is a lack of research to identify threats to automotive Ethernet-based networks. In this section, we introduce threats targeting automotive Ethernet-based IVNs. This section does not elucidate all possible threats to IVNs but emphasizes the need to identify attacks on automotive Ethernet. We assume that the attacker has access to a target IVN. \autoref{fig:topology} shows an example topology of automotive Ethernet-based IVN in addition to possible scenarios of how adversaries can invade the target IVN. Attackers can invade the IVN through physical access using the OBD-II port and through remote access via compromised infotainment devices. Once an attacker gains such access, (s)he is capable of (1) passive attacks and (2) active attacks.%

An attacker carries out a \textbf{passive attack} by monitoring inbound traffic on a particular node monitored by him/her. The goals of a passive attack include identifying available services, vehicle status, and nature of in-vehicle communications. For example, a passive attacker can identify the presence of a newly established AVTP stream from inbound control AVTPDUs. A passive attack is difficult to detect, as it does not affect any data within IVNs. Fortunately, the impact of a passive attack is relatively limited (compared with an attack on the CAN bus to which all nodes are connected). The reason is that in-vehicle switches transmit only broadcast packets to the monitoring node, whereas other packets reach only their designated destinations.%

\begin{table}[t]
\caption{Five active attacks on automotive Ethernet-based IVNs.}
\label{table:active_attacks}
\centering
\scriptsize{
\begin{tabular}{p{1.5cm}lp{3.5cm}lll}
\toprule
Attack & Target & Goal & Likelihood & Risk & Discriminability \\ \midrule
DoS attack           & ECU    & Out-of-communication status for a target ECU                  & High & High & Easy     \\ 
CAM table overflow   & Switch & Disable switching and broadcast all legitimate traffic within IVNs   & High & Low  & Easy     \\ 
Fuzzing              & ECU    & Execute an unexpected/hidden command on the target ECU & High & High & Easy    \\ 
Command injection & ECU    & Execute a command as intended by the attacker             & Low  & High & Moderate \\ 
Replay attack &
  ECU &
  Re-execute instructions executed at the time of traffic extraction &
  Middle &
  High &
  \textbf{\underline{Hard}} \\ \bottomrule
\end{tabular}
}
\end{table}

An \textbf{active attack} begins with a packet injection performed by the attacker. The attacker can inject arbitrary packets. Hence, injection attacks can be classified into several types depending on the attacker's purpose and the effects of the attack on IVNs. Here, we introduce five types of plausible active attacks (but further discovery and assessment are also needed):%

\begin{itemize}
    \item \BfPara{Denial of service (DoS) attack} An attacker may attempt a DoS attack in a specific ECU by flooding it with numerous VLAN-tagged packets. Automotive Ethernet is vulnerable to such an attack because it prioritizes network traffic and ensures a certain quality of service using a 3-bit \textit{PCP} field in the VLAN header. Thus, the packets injected by an attacker can override other legitimate packets transmitted to the target ECU. Specifically, the flooder initiates a burst transmission to the switch of the target ECU. This causes packet loss on the network level; other ECUs cannot transmit any packets to the target ECU during the DoS attack. However, the DoS attack does not affect other ECUs. Note that each stream AVTPDU contains a VLAN header with a value of 3 in the \textit{PCP} field (depicted in row 3 in \autoref{fig:header}). Consequently, an AVB listener may not receive an AVTP stream owing to the DoS attack.%
    
    \item \BfPara{Content-addressable memory (CAM) table overflow} A CAM attack exploits the MAC address learning process of in-vehicle switches. Specifically, the attacker floods the network with packets having a random source MAC address until the target switch cannot accept new MAC address-port pair entries. This occurs when the CAM table of the target switch is full. As a result, all inbound traffic is flooded to all ECUs connected to the target switch. However, this threat exists not only in automotive Ethernet switches but also in general switches \cite{Alabady2008DesignServer}. Commercial switches can be equipped with countermeasures for this attack \cite{vyncke2006layer2sec}. Nevertheless, we should ensure that in-vehicle switches are robust against CAM table overflow attacks, considering that this is a period of transition from the use of the traditional CAN bus to automotive Ethernet. If an in-vehicle switch is vulnerable to the CAM table overflow attack, the impact of passive attacks will be very high because a single monitoring node can dump all transmitted packets.%
    
    \item \BfPara{Fuzzing} In a fuzzing attack, random inputs are given to a target ECU in the form of a network packet. In this case, the attacker can scan active services, disrupt communications, and execute unexpected commands for a designated service. Furthermore, a fuzz packet can occasionally trigger an undocumented service mode (e.g., ECU firmware upgrade mode) or change parameters that are crucial for vehicle operation. The black-box fuzzing attack is easy to perform because it requires less effort to affect the target vehicle. Furthermore, a gray-box fuzzing attack conducted by an attacker with prior knowledge may cause a critical accident. Most of the traffic generated by a fuzzer contains a malformed payload, which makes such traffic easy to detect in an IVN implementation.%
    
    \item \BfPara{Command injection} This attack is performed by an attacker who knows the specification of a target vehicle in detail. The attacker injects well-crafted payloads to the IVN of the target vehicle. Then, the target vehicle executes commands as assigned. The nature of command injection depends on the protocol or application targeted by an attacker. For example, packet injection may take place over a varying time interval, depending on the type of attack. Rezvani \cite{rezvani2018hacking} discussed the working of an automotive Ethernet camera and demonstrated a command injection attack. He further exploited the \textit{end of image function} of the target camera to override a video stream in accordance with his commands. To detect a command injection attack, an IDS needs to recognize the context of the communication or know the signatures captured from the injected packets. The risk of command injection attacks is very high because the target vehicle executes commands as intended by the attacker.%
        
    \item \BfPara{Replay attack} A replay attack enables an attacker to input a few commands even without prior knowledge of the attack target. To execute a replay attack, the attacker must possess a pre-captured packet dump. By replaying the packet dump to the IVN, the attacker can make the target vehicle execute certain instructions (which were captured during the traffic extraction). With regard to an IDS, detecting a replay attack is straightforward because there will be two ongoing sessions during such an attack, which is uncommon. However, it is challenging to determine which session or packets are being replayed by the attacker (because the injected packets were also generated by legitimate ECUs).%
\end{itemize}

\autoref{table:active_attacks} summarizes five types of active attacks. We evaluate the \textit{likelihood}, \textit{risk}, and \textit{discriminability} of the attacks based on the attack characteristics discucsed previously. Here, we assign a high \textit{likelihood} when no prior knowledge or effort is required before the attack. We assign a high \textit{risk} if the target ECU can become out-of-order or can execute an attacker's command. Finally, \textit{discriminability} indicates whether the injected packet is distinct from normal traffic (the packets can be easily detected if the injected packet is distinct). DoS attacks are relatively easy to detect because such attacks involve fixed payloads. CAM table overflow and fuzzing attacks are also easy to detect because they involve the injection of randomly generated packets. The detection of a command injection attack requires appropriate knowledge; therefore, it involves moderate difficulty. In contrast, replay attacks are very difficult to detect because the replayed packets have already been used in the target vehicle. Therefore, it is necessary to develop an IDS particularly for replay attacks.%

%% file: paper/04_system_design.tex
\section{System design}
\label{sec:system_design}

\begin{figure}[t]
\centering
\includegraphics[width=\linewidth]{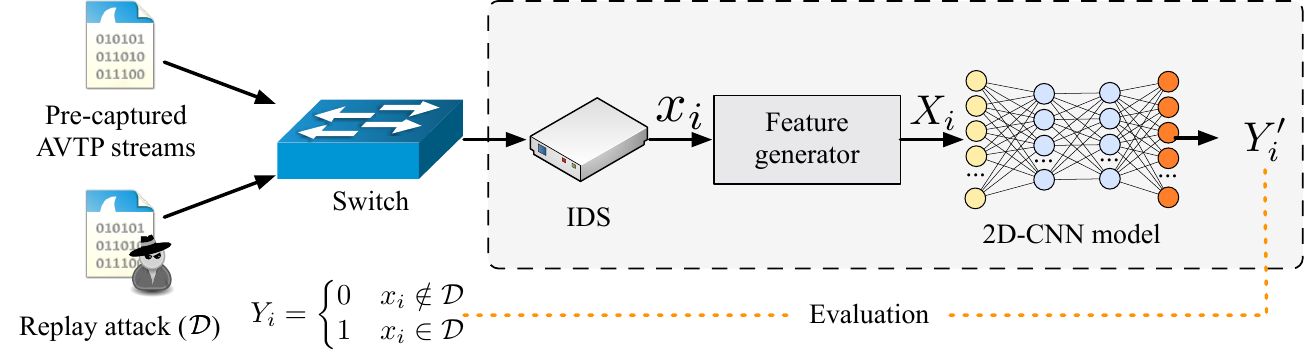}
\caption{Summary of our system design.}
\label{fig:topology_experiment}
\end{figure}

In this section, we propose an IDS designed to detect continuous replay attacks abusing stream AVTPDUs. In \autoref{subsec:adversary}, we demonstrate an example of the replay attack on automotive Ethernet. \autoref{fig:topology_experiment} depicts the proposed system design.

\BfPara{IDS deployment} We assume that the IDS should monitor consecutive AVTP stream to detect intrusions. To this end, we describe how an IDS can be deployed in a CAV. \autoref{fig:topology} shows that the IDS is connected to one of AVB listeners. Because AVTP supports one-to-many streams, IDS can observe AVTP streams through an AVB listener without affecting other in-vehicle systems.

\subsection{AVB packets}
\label{subsec:dataset}

Before we develop an intrusion detection method for replay attacks on automotive Ethernet, we need to gather automotive Ethernet packets. To this end, we first implemented a testbed using a BroadR-Reach physical network to capture AVB-related packets. We connected an AVB talker and an AVB listener to a BroadR-Reach switch. Then, we configured our BroadR-Reach switch to forward all of the inbound packets to a monitoring port. Hence, we accrued all of the communications between the AVB devices. 

Initially, the two devices synchronize through gPTP packets and send control AVTPDUs to establish a new session. After the connection is established, the AVB talker transmits one-way stream AVTPDUs. The stream AVTPDUs contain a compressed MPEG-2 TS live video stream from a USB camera directly connected to the AVB talker. The AVB listener plays the video by combining the received packets. All such communication processes are stored as one packet dump. We collected several packet dumps for the experiment (the pre-captured AVTP streams are shown in \autoref{fig:topology_experiment}).%

\subsection{Adversary}
\label{subsec:adversary}

\begin{figure}[t]
\centering
\subfloat[Live stream from the AVB talker -- \textit{``Must stop to prevent a traffic accident.''} ]{\frame{\includegraphics[width=0.24\linewidth]{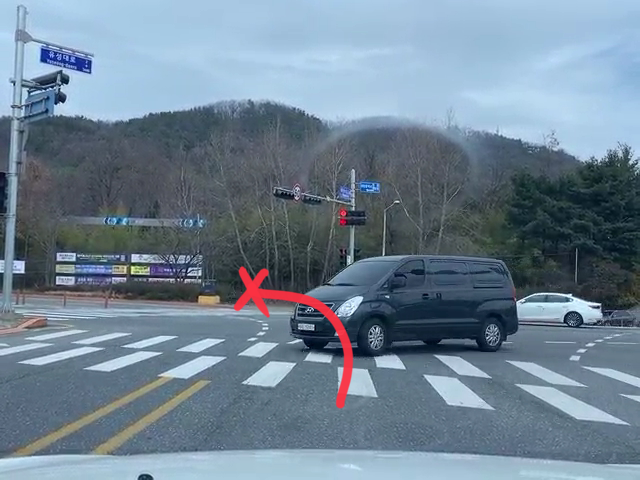}}%
\label{subfig:capture_original}}
\hfill
\subfloat[Injected stream from the adversary -- \textit{``Looks good to go.''}]{\frame{\includegraphics[width=0.24\linewidth]{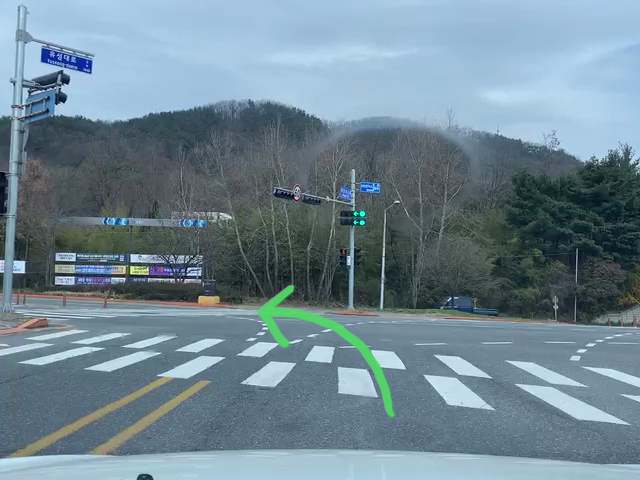}}%
\label{subfig:capture_injected}}
\hfill
\subfloat[Received footage at the AVB listener -- \textit{``The vehicle may depart due to ADAS's incorrect judgment based on the current input stream.''}]{\frame{\includegraphics[width=0.24\linewidth]{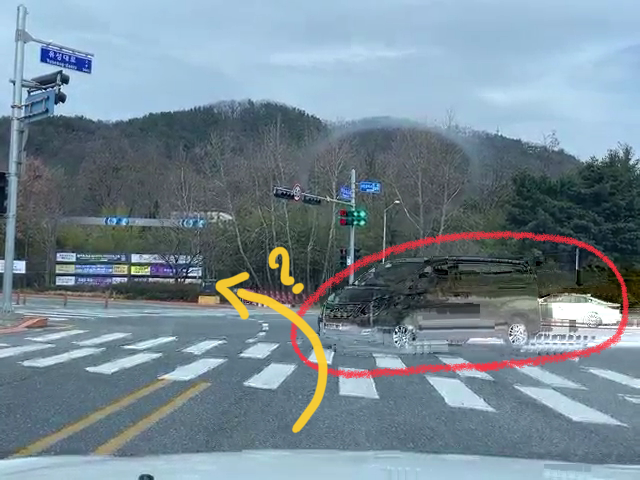}}\frame{\includegraphics[width=0.24\linewidth]{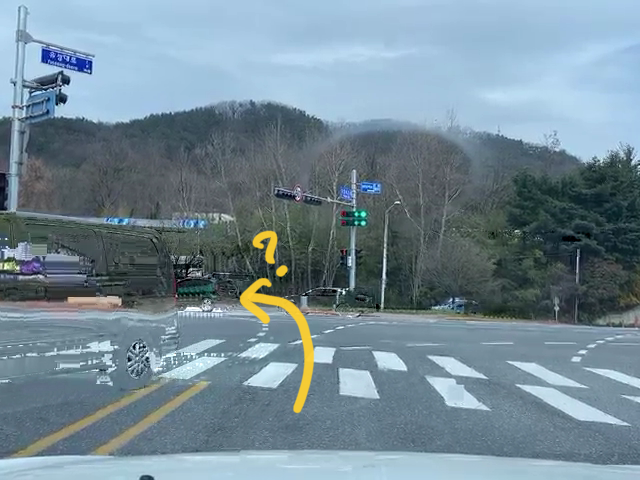}}%
\label{subfig:capture_result}}
\\
\caption{Demonstration of a replay attack (captured from our BroadR-Reach network). The terminal application plays a distorted video, as well as a little bit of the residual video; however, it mostly plays the scene intended by the attacker, as depicted in (c). Assuming that an autonomous driving system is connected to the AVB listener, a traffic accident is impending in this example scenario.}
\label{fig:screenshot}
\end{figure}

We suppose that an attacker injects arbitrary stream AVTPDUs into the IVN. The goal of the attacker is to output a single \textit{video frame}, at a terminal application connected to the AVB listener, by injecting previously generated AVTPDUs during a certain period. To demonstrate the attack, we extract 36 continuous stream AVTPDUs from one of our AVB datasets; the extracted AVTPDUs constitute one \textit{video frame}. Then, the attacker performs a \textbf{\textit{replay attack}} by sending the 36 stream AVTPDUs repeatedly.%

To illustrate the effect of the replay attack, \autoref{fig:screenshot} shows the data collected from our testbed during this attack. As shown in \autoref{subfig:capture_result}, the adversary successfully compromises the video stream on the AVB listener side. The terminal application plays corrupted video footage when the replay attack occurs (because the AVB listener receives stream AVTPDUs from two talkers). The attack can have a critical impact if the terminal application is the ADAS; a CAV has no choice but to be blinded or make a wrong decision. Therefore, it is important to detect injection attacks and replay attacks.%

\subsection{Payload observations}

\begin{figure}
\centering
\subfloat[Benign stream -- 438 bytes of total payload]{\includegraphics[width=\linewidth]{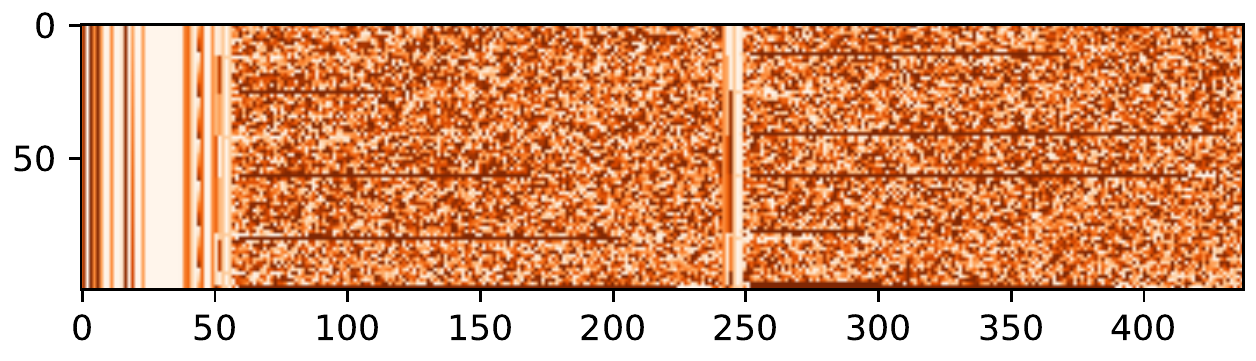}%
\label{subfig:visualized_stream_all}}
\\
\subfloat[Benign stream -- first 58 bytes]{\includegraphics[width=0.45\linewidth]{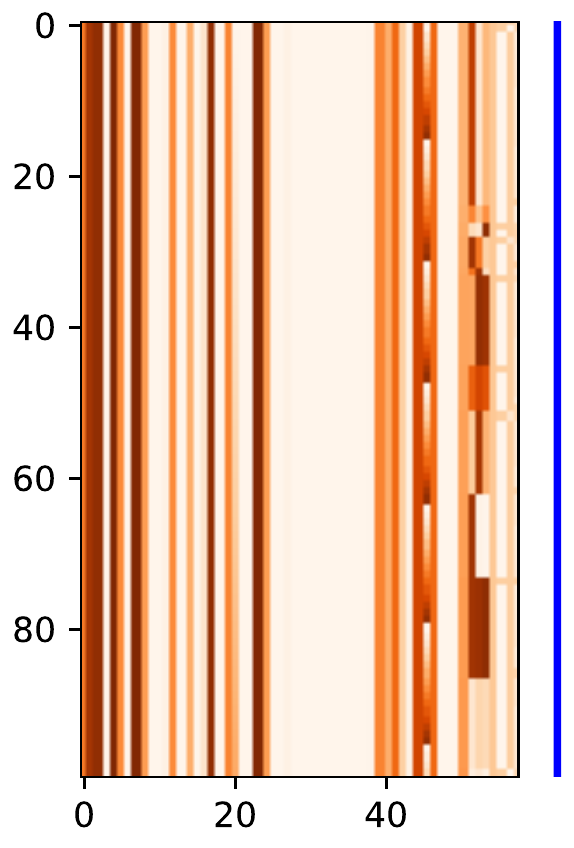}%
\label{subfig:visualized_stream_58_benign}}
\hfill
\subfloat[Injected stream -- first 58 bytes]{\includegraphics[width=0.45\linewidth]{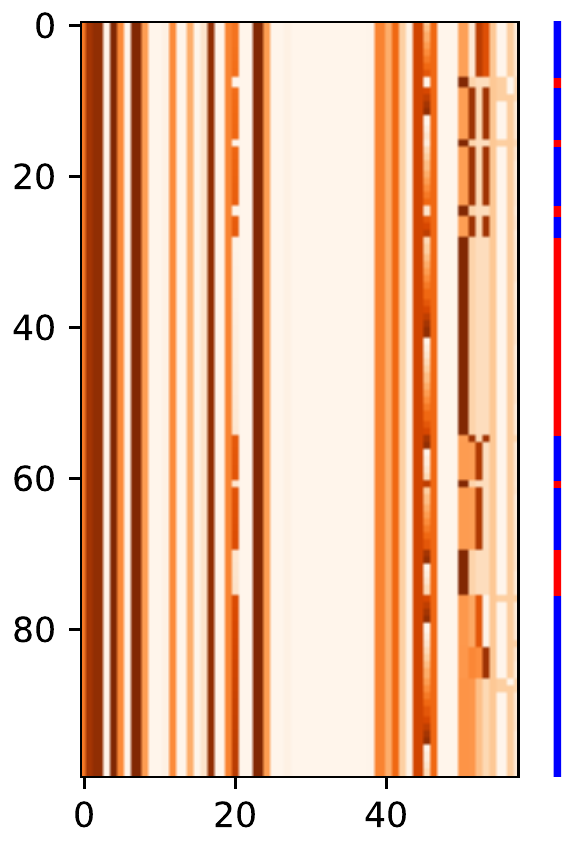}%
\label{subfig:visualized_stream_58_injected}}
\caption{Visualization of 100 continuous stream AVTPDU payloads. The vertical bars in (b) and in (c) represent whether the stream is benign (blue) or injected (red).}
\label{fig:visualized_stream}
\end{figure}

In this section, we present some observations of stream AVTPDU payloads to explain their characteristics. We also elucidate the motivation for our design of the packet generator.

\autoref{fig:visualized_stream} shows the visualization of 100 continuous-stream AVTPDU payloads captured from our experiment testbed. Each pixel represents a byte value: light and dark cells indicate low and high values, respectively. Values are given in the order of incoming AVTPDUs (from the top row to the bottom row). Each stream AVTPDU that we captured has the same size (438 bytes).

We found some interesting aspects in the first 58 bytes of the stream AVTPDUs \autoref{subfig:visualized_stream_58_benign}. First, each header of the stream AVTPDUs, shown in \autoref{fig:header}, has a fixed value, except for the \texttt{sequence num}, \texttt{DBC}, and \texttt{Payload} fields. Second, the values of the \texttt{sequence num} and \texttt{DBC} fields sequentially increase by 01 h and 10 h, respectively, and are set to 00 h when it overflows. Third, values in the offset range of 51--58 (the first eight bytes of the \texttt{Payload} field) change occasionally. This pattern repeats every time a new video frame is received.

In contrast, we found no meaningful pattern in the byte offset range of 59-438, although a light vertical line (offset 248) and a dark horizontal line were observed. The offset range of 59-438 is part of the MPEG-2 TS payloads carried by the AVTPDUs.

\autoref{subfig:visualized_stream_58_injected} visualizes the payloads during a replay attack. The injected frames can be distinguished using the vertical line on the right. Conspicuous grid-like patterns can be observed on the aforementioned fields; the patterns correspond to the alternation of normal and injected stream AVTPDUs. Such a pattern means that the fields representing the sequence or the context in the payload are fragmented. However, the replay attack may seem like a normal flow when the adversary dominates the network traffic (see near the 40th packet). This makes it difficult to detect replayed packets.%

There is no change in the Ethernet header or the VLAN header. Malformed stream AVTPDUs injected by a weak adversary should look conspicuous and awkward in the benign stream in automotive Ethernet-based networks. For example, we can easily detect a DoS attack based on a modified VLAN header. An IDS can quickly detect such injections with substantially less effort; an example of this is rule-based intrusion detection that observes the modification of static payloads.%

\subsection{Feature generator}

\begin{figure}
\centering
\includegraphics[width=\linewidth]{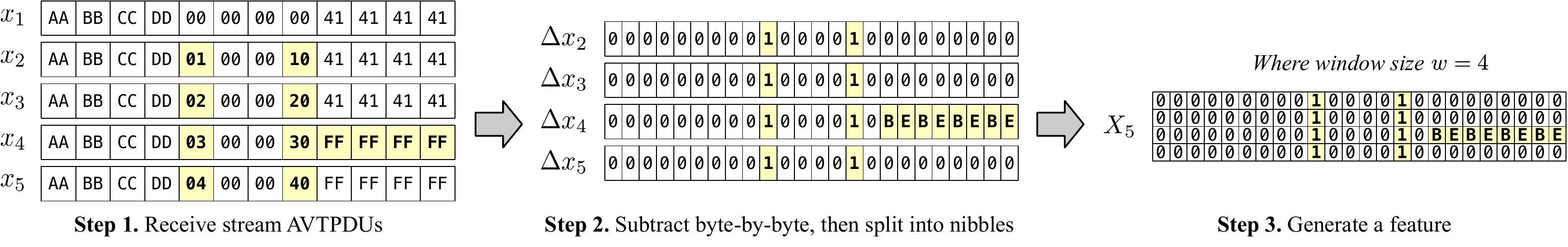}
\caption{Proposed feature generation process}
\label{fig:feature_generation}
\end{figure}

Next, we design the feature generator so that the proposed CNN model can distinguish and classify the changes in the payload after a packet injection. The feature generator receives a stream AVTPDU from a BroadR-Reach network and generates a feature to be used as an input of  the CNN. Based on our observations, we decided to focus on the first 58 bytes of the $i$-th arriving stream AVTPDU. Moreover, the feature generator uses a window size $w \in \mathbb{N} \mid w \geq 4$; it aggregates the recently arrived traffic using $w$ and then produces a two-dimensional vector from it.

Let $x_i$ be a one-dimensional vector containing $j$-bytes of $i$-th stream AVTPDU, which is given as $x_i=(x_{i,1}, x_{i,2}, \ldots, x_{i,(j-1)}, x_{i,58})$ where $x_{i,j} \in \mathbb{N} \mid 0 \leq x_{i,j} < 256$. Thus, vector $x_i$ includes all fields and only the first 8 bytes of \texttt{Payload} field. We set $j=58$ because we consider the first 58 bytes of each stream AVTPDU payload. Considering two stream AVTPDUs $x_{i-1}$ and $x_i$, let $\Delta x_i$ be the state change of the $i$-th stream AVTPDU, which is derived as follows:
\begin{equation}
\begin{aligned}
\Delta x_i  \equiv & (x_i - x_{i-1}) \mod{2^8} \\
            \equiv & (x_{i,1}-x_{(i-1),1}, \ldots, x_{i,58}-x_{(i-1),58}) \mod{2^8} \\
            \equiv & (\Delta x_{i,1}, \Delta x_{i,2}, \Delta x_{i,3}, \ldots, \Delta x_{i,(j-1)}, \Delta x_{i,58}) \\
            = & (u_{i,1}, v_{i,1}, u_{i,2}, v_{i,2} \ldots, u_{i,58}, v_{i,58})
\end{aligned}
\end{equation}

where $\Delta x_{i,j} = u_{i,j}2^{4}+v_{i,j}$ and $u_{i,j},v_{i,j} \in \mathbb{N} \mid 0 \leq u_{i,j},v_{i,j} < 16$.  Finally, a $w \times 2j $ sized two-dimensional vector $X_i$ is defined as

\begin{equation}
\begin{aligned}
X_i = &
   \begin{pmatrix}
   \Delta x_{i-w+1} \\
   \Delta x_{i-w+2} \\
   \vdots \\
   \Delta x_{i} 
   \end{pmatrix} 
\\ = &
    \begin{pmatrix}
   u_{(i-w+1),1} & v_{(i-w+1),1} & \cdots & v_{(i-w+1),58} \\
   u_{(i-w+2),1} & v_{(i-w+2),1} & \cdots & v_{(i-w+2),58} \\
   \vdots  & \vdots  & \ddots & \vdots  \\
   u_{i,1} & v_{i,1} & \cdots & v_{i,58} 
   \end{pmatrix}
\end{aligned}
\end{equation}

This expression gives the window size corresponding to a recent stream change.

\autoref{fig:feature_generation} illustrates the process of feature generation. In summary, the feature generator returns $X_i$ on the arrival of the $i$-th stream AVTPDU once the window is full (i.e., $i > w$). The feature generator feeds each $X_i$ to the proposed CNN model. $Y_i \in \{0, 1\}$ is employed to identify whether the $i$-th stream AVTPDU is benign or injected; this data is then used for training our CNN model.

\subsection{CNN-based intrusion detection model}

\begin{table}[t]
\centering
\caption{Structure of the convolutional neural network and the output dimensions (where $w = 44$)}
\label{table:layers}
\scriptsize{
\begin{tabular}{@{}llllp{2.7cm}}
\toprule
Role & Layer name & Data shape & Activation & Hyperparameters \\ \midrule
Input & Input ($X_i$) & $44\times116\times1$ & --- & --- \\ \midrule 
\multirow{6}{*}{\begin{tabular}[l]{@{}l@{}}Feature\\ extraction\end{tabular}} & Conv2D\_1 & $44\times116\times32$ & ReLU & filter\_size=$5\times5\times32$, stride=$1\times1$, ``same'' padding, L2 reg. \\  
 & BatchNormalization\_1 & $44\times116\times32$ & --- & momentum=0.99,
    epsilon=0.001 \\  
 & MaxPool\_1 & $22\times58\times32$ & --- & pool\_size=$2\times2$\\ 
 & Conv2D\_2 & $22\times58\times64$ & ReLU & filter\_size=$5\times5\times64$, stride=$1\times1$, ``same'' padding, L2 reg. \\  
 & BatchNormalization\_2 & $22\times58\times64$ & --- & momentum=0.99,
    epsilon=0.001 \\  
 & MaxPool\_2 & $11\times29\times64$ & --- & pool\_size=$2\times2$\\ \midrule 
\multirow{5}{*}{\begin{tabular}[l]{@{}l@{}}Binary\\ classification\end{tabular}} & Flatten & $20416$ & --- & --- \\  
 & Dropout\_1 & $20416$ & --- & dropout\_ratio=0.3 \\  
 & Dense\_1 & $64$ & ReLU & --- \\  
 & Dropout\_2 & $64$ & --- & dropout\_ratio=0.3 \\  
 & Dense\_2/Output ($Y_{i}'$) & $1$ & Sigmoid & --- \\ \bottomrule
\end{tabular}
}
\end{table}

Our CNN model comprises an input layer, two hidden sub-layers, and two dense layers. \autoref{table:layers} details the proposed CNN architecture and hyperparameters. Window size $w$ must be a multiple of 4 owing to size reduction in two max-pooling layers. To express the dimension of each layer, we use window size $w=44$, which is the optimal value for our dataset (see \autoref{subsec:windowsize}). Thus, the input layer accepts $44\times116\times1$-sized data.

Each hidden sub-layer contains a convolution layer, a batch normalization layer, and a max-pooling layer. The \textit{batch normalization} layers prevent overfitting and accelerate the training phase \cite{ioffe15batch}. In the two \textit{Conv2D} layers, we use L2 regularization as the kernel regularizer, which limits the values of the weights. We deploy two \textit{dropout} layers between fully connected layers, which also prevents overfitting \cite{JMLR:v15:srivastava14a}. The rectified linear unit (ReLU) activation function gives non-linearity to the neural network and is widely used in CNN architectures to learn complex features in input data.

The last dense layer returns $Y_{i}'$ through a \textit{sigmoid} activation function, where $Y_{i}' \in \mathbb{R} \mid 0\leq Y_{i}'\leq 1$. We consider output $Y_{i}'$ as the probability of $X_i$ being injected . In the training phase, the binary cross-entropy loss function compares $Y_{i}'$ to $Y_{i}$. Our CNN model is trained using the Adam optimizer \cite{Kingma2014}, with a learning rate of 0.001.

%% file: paper/05_experiment_result.tex
\section{Experiment}
\label{sec:experiment}
\subsection{Environment}
\label{subsec:environment}
In the experiments, we use the Python library Keras \cite{chollet2015keras} to implement our 2D-CNN model. We train and evaluate our model using Google Colaboratory on an NVIDIA Tesla P100 GPU.

We performed four packet captures of attack-free stream AVTPDUs from our BroadR-Reach network in the PCAP format for 70 minutes in total. The first two packet captures are were performed indoors (in our laboratory) for 25 min, and the last two packet captures were performed for 45 min in a vehicle driving on the road. We noticed that the packet transmission rates were different in both environments. A possible reason is that the camera connected to the AVB talker was capturing a more dynamic scene during driving. For instance, the average interval between packets is 3,157 $\mu$s indoors and 1,735 $\mu$s for a moving vehicle.

Each packet capture corresponds to a single AVB session, including broadcast messages for node discovery, establishment of an AVB session, video transmission over stream AVTPDUs, and disconnection of the AVB session.

\subsection{AVTP intrusion dataset}
To demonstrate a replay attack, we continuously inject a \textit{video frame} (the 36 stream AVTPDUs discussed in \autoref{subsec:adversary}) and recapture the four packets. Note that we waited for 120 seconds before the attack to obtain some data corresponding to the attack-free scenario, which allowed our 2D-CNN model to learn characteristics corresponding to the benign status as well. Then, we turned the four captured packets into AVTP intrusion datasets using the feature generator. All packets except for stream AVTPDUs were excluded during the feature generation.

We denote the AVTP intrusion datasets as \dataset1 and \dataset2. We selected \dataset1  as the training/validation set. \dataset1 was used to choose proper values for $w$ and for the hyperparameters of our 2D-CNN model. We used \dataset2 as the test set. Note that the packet collection environments of both datasets are independent. Thus, there can be no potential over-fitting caused by unrecognized payloads. \dataset1 has 446,372 benign $X_i$s and 196,894 injected $X_i$s. \dataset2 has 1,494,257 benign $X_i$s and 376,236 injected $X_i$s.

\subsection{Evaluation metrics}
We use Accuracy, Precision, and Recall as performance metrics. The \textit{True Positive (TP)} and \textit{True Negative (TN)} are a number of AVTPDUs correctly classified as benign and injected, respectively. On the other hand, the \textit{False Positive (FP)} and \textit{False Negative (FN)} are a number of AVTPDUs incorrectly classified as benign and injected, respectively. Then, the performance metrics are defined as:

We use accuracy, precision, and recall as performance metrics.\textit{True positive (TP)} and \textit{true negatives (TN)} are the numbers of AVTPDUs correctly classified as benign and injected, respectively. \textit{False positives (FP)} and \textit{false negatives (FN)} are the numbers of AVTPDUs incorrectly classified as benign and injected, respectively. Thus, the performance metrics are defined as follows:

\begin{equation}
\begin{aligned}
\rm{Accuracy} = & \rm{(TP+TN)/(TP+TN+FP+FN)} \\
\rm{Precision} = & \rm{TP/(TP+FP)} \\
\rm{Recall} = & \rm{TP/(TP+FN)} \\
\end{aligned}
\end{equation}

Moreover, the F1-score is the harmonic mean of precision and recall, which is sometimes more useful than accuracy, especially when the class distribution is imbalanced. The F1-score is calculated as follows:

\begin{equation}
\rm{F1\operatorname{-}score} = 2 \times \rm{Precision} \times \rm{Recall}/(\rm{Precision}+\rm{Recall})
\end{equation}

We also used the receiver operating characteristic (ROC) curve and the area under the ROC curve (AUC) to demonstrate the accuracy of our model with regard to the classification of $X_i$s. The ROC curve is a common method to measure the performance of binary classifiers. The ROC curve shows changes in the true-positive rate (TPR) and false-positive rate (FPR) according to the threshold change for $Y_{i}'$. The TPR and FPR are calculated as follows:

\begin{equation}
\begin{aligned}
\rm{TPR} = & \rm{Recall} = \rm{TP/(TP+FN)} \\
\rm{FPR} = & \rm{FP/(TN+FP)} \\
\end{aligned}
\end{equation}

\subsection{Choosing an optimal window size}
\label{subsec:windowsize}
To find an optimal $w$, we created 20 \dataset1 sets for $w=(4, 8, 12, ..., 76, 80)$ using the feature generator. Then, we built and trained 20 2D-CNN models with different input shapes. We randomly assigned 80\% of $X_i$s to the training set and 20\% of $X_i$s to the test set for each \dataset1.

\autoref{table:experiment_window_size} presents the performance of the model with respect to $w$. The best result in each column is highlighted. When a smaller window size is employed, the training requires less time. However, the classification performance is poor for small window sizes. Conversely, one might conclude that a large $w$  will be advantageous because a wide window allows the 2D-CNN model to capture more information from the input data. However, the largest $w$ in our experiment does not bring the best result. Instead, the best performance was obtained for $w = 44$ for the test set and $w = 60$ for the training set. Therefore, we chose 44 as the optimal window size because the model is more robust to the test set.

\begin{table}
\caption{Model performance on our dataset depending on window size $w$. The best outcome in each column is highlighted. The result shows that the optimal value of $w$ is 44 or 60 (instead of the highest value).}
\label{table:experiment_window_size}
\centering
\begin{tabular}{llllll}
\toprule
\textbf{$w$} & \begin{tabular}[l]{@{}l@{}}Training time\\ ($\mu$s/sample)\end{tabular} & \begin{tabular}[l]{@{}l@{}}Train\\ loss\end{tabular} & \begin{tabular}[l]{@{}l@{}}Train\\ accuracy\end{tabular} & \begin{tabular}[l]{@{}l@{}}Test\\ accuracy\end{tabular} & \begin{tabular}[l]{@{}l@{}}Test\\ F1-score\end{tabular} \\ \midrule
4 & \textbf{\underline{81}} & 0.4454 & 0.7578 & 0.7603 & 0.7603 \\ 
8 & 96 & 0.2996 & 0.8249 & 0.8303 & 0.8303 \\ 
12 & 104 & 0.1981 & 0.9101 & 0.9126 & 0.9126 \\ 
16 & 115 & 0.1225 & 0.9595 & 0.9633 & 0.9633 \\ 
20 & 134 & 0.0915 & 0.9771 & 0.9781 & 0.9781 \\ 
24 & 146 & 0.0666 & 0.9856 & 0.9867 & 0.9867 \\ 
28 & 153 & 0.0606 & 0.9878 & 0.9928 & 0.9928 \\ 
32 & 166 & 0.0575 & 0.9889 & 0.9903 & 0.9903 \\ 
36 & 180 & 0.0523 & 0.9894 & 0.9897 & 0.9897 \\ 
40 & 198 & 0.0741 & 0.9851 & 0.9929 & 0.9929 \\ 
44 & 206 & 0.0520 & 0.9891 & \textbf{\underline{0.9954}} & \textbf{\underline{0.9954}} \\ 
48 & 216 & 0.0574 & 0.9871 & 0.9931 & 0.9931 \\ 
52 & 228 & 0.0513 & 0.9893 & 0.9939 & 0.9939 \\ 
56 & 238 & 0.0551 & 0.9884 & 0.9868 & 0.9868 \\ 
60 & 267 & \textbf{\underline{0.0493}} & \textbf{\underline{0.9903}} & 0.9916 & 0.9916 \\ 
64 & 275 & 0.0554 & 0.9869 & 0.9109 & 0.9109 \\ 
68 & 269 & 0.0682 & 0.9853 & 0.9929 & 0.9929 \\ 
72 & 276 & 0.0622 & 0.9863 & 0.9873 & 0.9873 \\ 
76 & 296 & 0.0687 & 0.9842 & 0.9911 & 0.9911 \\ 
80 & 307 & 0.0566 & 0.9873 & 0.9877 & 0.9877 \\ \bottomrule
\end{tabular}
\end{table}

\subsection{Experiment results}
We performed a five-fold cross-validation on \dataset1 to ensure that our 2D-CNN model is sufficient to detect intrusions and that the hyperparameters are properly determined. In each cross-validation, 80\% and 20\% of the samples are randomly selected as the training and validation sets, respectively. Each input is used once for validation. In each cross-validation, the training set and the validation set have the same proportion of benign/injected labels. After the cross-validation, we obtain five trained 2D-CNN models. 

We tuned the batch size to 64 and epoch to 30 for training our model. To evaluate the final performance of our method, we performed five experiments using each model on \dataset2.

\begin{table*}[t]
\centering
\caption{Classification results of five-fold cross-validation using dataset $\mathcal{D}_{\rm{indoors}}$}
\label{table:result_5cv}
\begin{tabular}{llllll}
\toprule
Fold & Accuracy & Precision & Recall & F1-score & AUC \\ \midrule
1     & 0.9958 & 0.9916 & 0.9947 & 0.9932 & 0.9998 \\
2     & 0.9967 & 0.9937 & 0.9958 & 0.9947 & 0.9999 \\
3     & 0.9945 & 0.9895 & 0.9927 & 0.9911 & 0.9997 \\
4     & 0.9946 & 0.9888 & 0.9937 & 0.9912 & 0.9998 \\
5     & 0.9958 & 0.9931 & 0.9932 & 0.9932 & 0.9998 \\ \midrule
Total & 0.9955 & 0.9913 & 0.9940 & 0.9927 & 0.9997 \\
\bottomrule
\end{tabular}
\end{table*}

\autoref{table:result_5cv} shows the classification results obtained for the five-fold cross-validation. As our dataset is slightly imbalanced, we should focus on the F1-score (although the accuracy is better). The experimental results show that the model exhibited outstanding performance, with F1-scores from 0.9911 to 0.9947. Thus, the model can classify almost all stream AVTPDUs correctly. The high recall implies that the proposed model is very sensitive with regard to the detection of almost all injected stream AVTPDUs, even if the attacker sends AVTP traffic previously generated in the target automotive Ethernet-based network. We listed  the validation results obtained for each cross-validation and created the ``total'' row in \autoref{table:result_5cv}. In conclusion, we achieved a satisfactory performance of 0.99 or higher for all evaluation indicators for \dataset1.

Notably, there is no significant difference in performance metrics among the five cross-validations. Hence, the training process is stable with the chosen hyperparameters. Moreover, our 2D-CNN model converges after the training phase regardless of input data. In other words, the experimental results obtained using \dataset1 demonstrate that our feature generator, 2D-CNN model, and hyperparameters can effectively detect AVTP replay attacks.

\begin{table}
\centering
\caption{Test results using dataset $\mathcal{D}_{\rm{driving}}$}
\label{table:result_test}
\begin{tabular}{llllll}
\toprule
Model & Accuracy & Precision & Recall & F1-score & AUC \\ \midrule
1     & 0.9953 & 0.9821 & 0.9949 & 0.9885 & 0.9996 \\ 
2     & 0.9930 & 0.9677 & 0.9990 & 0.9831 & 0.9992 \\ 
3     & 0.9877 & 0.9439 & 0.9984 & 0.9704 & 0.9979 \\ 
4     & 0.9923 & 0.9650 & 0.9984 & 0.9814 & 0.9996 \\ 
5     & 0.9913 & 0.9600 & 0.9988 & 0.9790 & 0.9989 \\ \bottomrule
\end{tabular}
\end{table}

\begin{figure}[!t]
\centering
\includegraphics[width=0.75\linewidth]{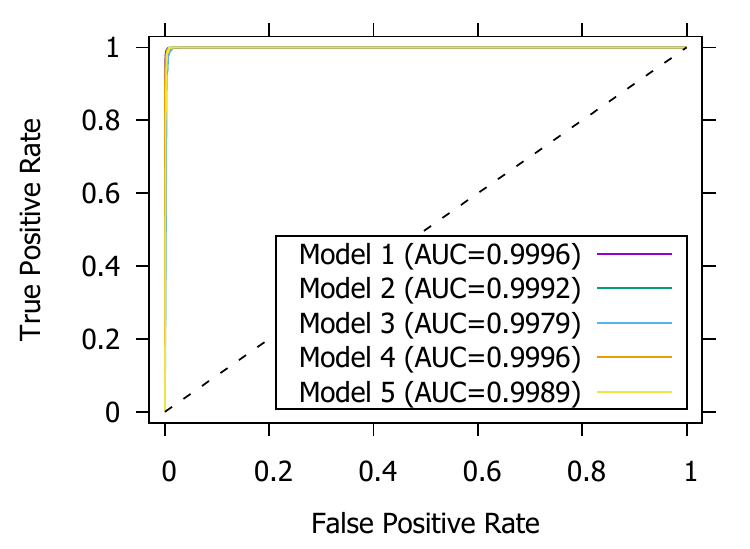}
\caption{ROC curve and AUC scores using test set $\mathcal{D}_{\rm{driving}}$}
\label{fig:roc_test}
\end{figure}

\autoref{table:result_test} shows the test results obtained using \dataset2. We performed five tests using the five models that were trained during the cross-validation. We found that the overall performance was slightly reduced compared to that observed for the cross-validation test. This is mostly due to a decrease in precision, meaning that there are considerable \textit{FP} cases in \dataset2. Interestingly, we confirmed that the recall had increased compared with that obtained for the validation result. Thus, we can conclude that our intrusion detection model (1) well distinguishes replayed stream AVTPDUs that are generally difficult to identify; however, (2) it may entail false-alarm fatigue during the attack-free period.

\autoref{fig:roc_test} shows that the AUCs are high although there was some performance reduction when the test set was employed. The ideal ROC curves and the high AUCs indicate that the 2D-CNN model well separates the input into benign and injected classes. It means that we can reduce the \textit{FP} cases, thereby improving the overall performance by adjusting the threshold for evaluating $Y_{i}'$.

\begin{figure}
\centering
  \subfloat[True negatives\label{fig:tn}]{\includegraphics[width=0.5\linewidth]{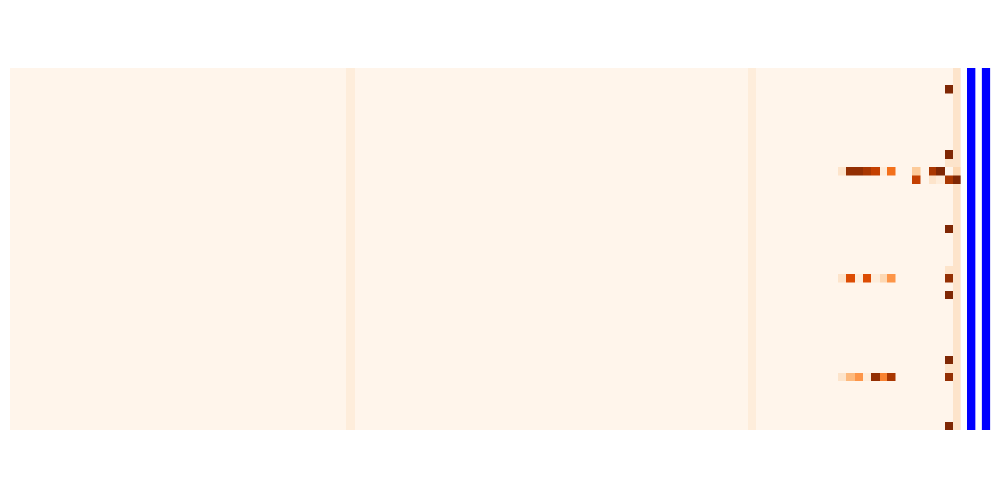}} \hfill
  \subfloat[False positives\label{fig:fp}]{\includegraphics[width=0.5\linewidth]{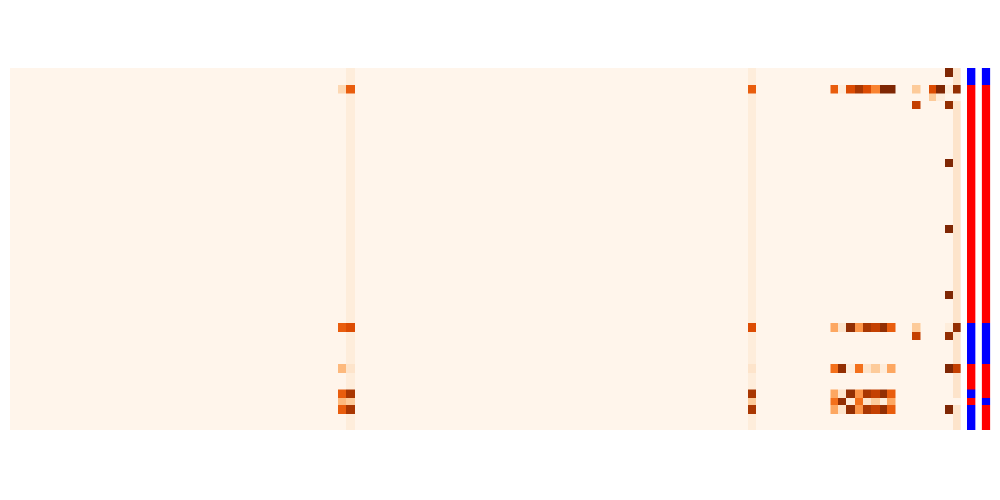}} \\
  \subfloat[False negatives\label{fig:fn}]{\includegraphics[width=0.5\linewidth]{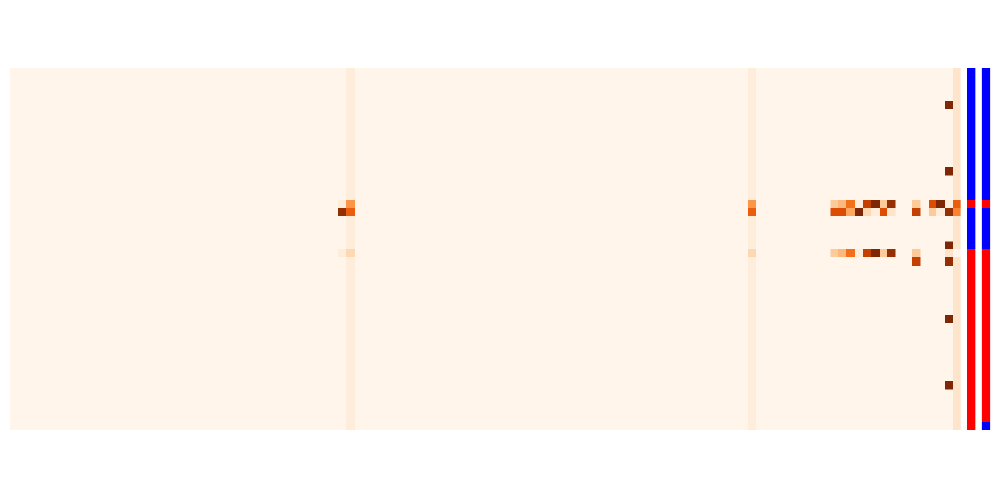}} \hfill
  \subfloat[True positives\label{fig:tp}]{\includegraphics[width=0.5\linewidth]{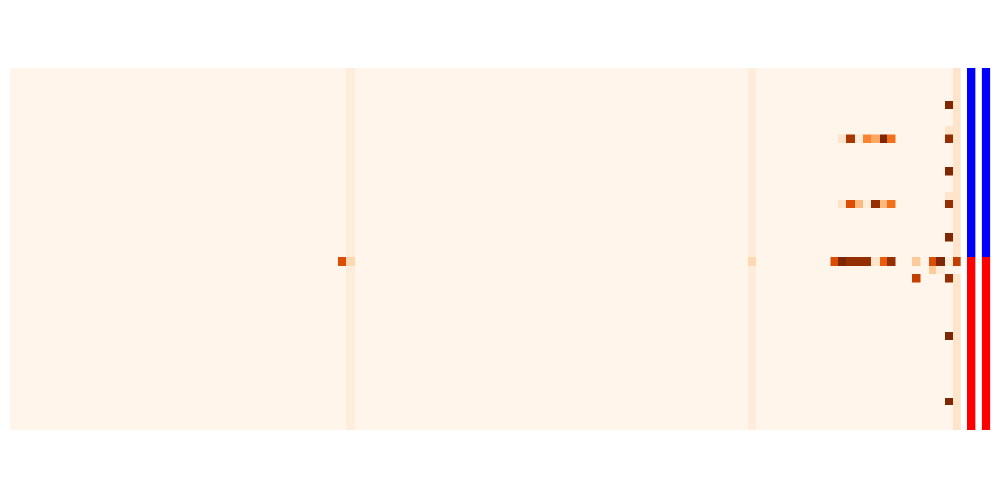}} \\
\caption{Visualization of four randomly selected samples ($X_i$s). Two vertical lines represent labels (left, $Y_{i-w+1}, ..., Y_{i}$) and predicted labels (right, $Y'_{i-w+1}, ..., Y'_{i}$). Blue and red dots mean that the traffic is (predicted as) ``benign'' and ``injected,'' respectively.} 
\label{fig:samples}
\end{figure}
  
To illustrate the input data and the classification results, we visualize four randomly selected $X_i$s from our dataset in \autoref{fig:samples}. Each dot expresses the extent of the state change of a nibble in the same position for adjacent stream AVTPDUs using brightness. The rightmost vertical bar represents $Y_{i-w+1}, ..., Y_{i}$ from top to bottom, in which blue and red dots mean ``benign'' (0) and ``injected'' (1), respectively.

\autoref{fig:samples}\subref{fig:tn} shows the typical state of stream AVTPDUs under the attack-free situation. It appears to have four \textit{video frames} because there are three horizontal patterns in the MPEG2-TS header.

\autoref{fig:samples}\subref{fig:tp} shows the attacker dominating the AVTP stream. Except for the first injected packet, all injected packets appear to form a normal stream. We can observe that our 2D-CNN model successfully infers all intrusions correctly.

When a dominant injection continues for a substantially long period, e.g., when more than half of the window size corresponds to injection, the injected packets are sometimes misclassified (see the bottom of \autoref{fig:samples}\subref{fig:fn}) because the features appear similar to those in the attack-free state. This is because the replay attack was conducted using the traffic generated in the target automotive Ethernet. Nevertheless, our model correctly identifies most of the continuous packet injections.

One of the weaknesses of our model, found in \autoref{fig:samples}\subref{fig:fp}, is that a single misclassification causes short-term continuous errors. In the visualization, we can observe five continuous classification errors.

\subsection{Toward real-time detection}

\begin{table*}[h]
\caption{Average inference time per sample for various devices}
\label{table:inference_time}
\centering
\begin{tabular*}{\linewidth}{llr}
\toprule
\textbf{Host} & \textbf{Processor} & \multicolumn{1}{l}{\textbf{\begin{tabular}[c]{@{}l@{}}Time \\ ($\mu$s/sample)\end{tabular}}} \\ \midrule
Google Colab & (GPU) NVIDIA Tesla P100  & 83 \\ 
Macintosh & (CPU) Intel Core i7-7700K  & 787 \\ 
Jetson TX2 & (GPU) NVIDIA Pascal (256 CUDAs) & 982 \\ 
Raspberry Pi 3 & (CPU) ARM Cortex-A53 & 35,000\\ \bottomrule
\end{tabular*}
\end{table*}

From a practical perspective, we need to consider not only the classification performance but also the inference time to determine whether the model is suitable for real-time detection in CAVs. To this end, we measured the average inference time per sample using four devices. \autoref{table:inference_time} shows the results. For real-time detection, the inference time per sample must be less than or equal to the packet occurrence time that we noted earlier in \autoref{subsec:environment}, i.e., 1,735 $\mu$s. We considered the threshold to be 1,000 $\mu$s, considering packet injections by an attacker.

In Google Colaboratory, our model takes only 83 $\mu$s to infer one sample with GPU acceleration, which is substantially shorter than the average packet interval. However, this result, obtained using a cloud artificial intelligence platform, does not reflect the situation in actual CAVs, because in-vehicle IDSs may not have such a powerful GPU. Therefore, we measured the inference time using a PC once again. Although the detection time increased substantially, it can be seen that real-time detection is possible using a CPU.

\begin{figure}[!t]
\centering
\includegraphics[width=\linewidth]{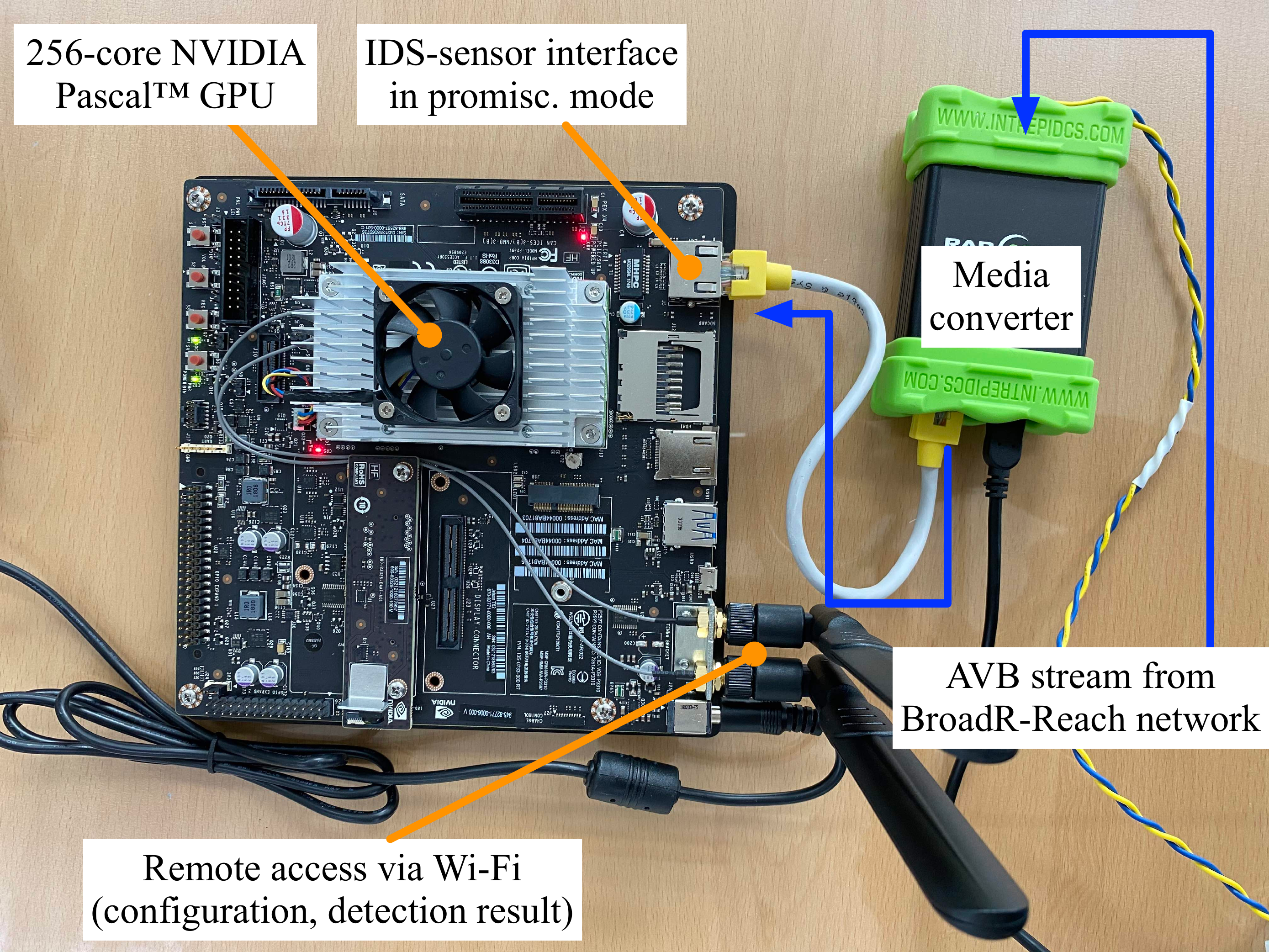}
\caption{Implementation of the real-time IDS designed to
identify injected stream AVTPDUs in a BroadR-Reach network}
\label{fig:physical_testbed}
\end{figure}

We further conducted the experiment on two embedded computers. We used a common device, Raspberry Pi 3, to show that the proposed model can function in a low-power computing environment. We successfully tested our approach on Raspberry Pi 3. However, the inference required a long time; hence, it was unsuitable for real-time detection. Next, we implemented an IDS for automotive Ethernet using an embedded GPU and a BroadR-Reach media converter.  \autoref{fig:physical_testbed} shows the implemented IDS connected to our testbed via BroadR-Reach. We used an NVIDIA Jetson TX2 and an Intrepid RAD-Moon media converter. The proposed model requires approximately 982 $\mu$s to process one sample in this case. According to these results, our CNN model is suitable for real-time detection on GPU-enabled embedded computers.

In general, computation complexity and classification accuracy are in a trade-off relationship. However, it is required to reduce computation and inference time if the in-vehicle system provides less computational performance or an IDS runs on an embedded system along with multiple applications, such as perception, decision making, and on-board diagnostics. To reduce inference time, model optimization needs to be done with resizing input size, model compression, or quantization. In \autoref{subsec:windowsize}, we confirmed that the smaller window size we use, the shorter the model's training time. For instance, changing window size from 44 to 28 decreases 25\% of training time and increases small misclassification error (about 0.26\% of test F1-score), which is endurable. The reduction in learning time applies equally to inference time. To compress or quantize the proposed model, additional hyperparameters need to be carefully considered. This requires a detailed study and is left for future work.

%% file: paper/06_discussion.tex
\section{Discussion}
\label{sec:discussion}

\subsection{Dataset}
During the experiment, we carefully created the automotive Ethernet intrusion datasets instead of conducting online learning. This helps us reproduce the experimental results and also lets other researchers study their own intrusion detection methods for automotive Ethernet. The datasets are recorded in the PCAP file format and, therefore, can be viewed using prevalent programming libraries and packet analyzers (such as Wireshark). As previously discussed, a replay attack was used as the intrusion technique. In this case, analysts without prior knowledge cannot distinguish the states of the packets as benign or injected. To address this, we have labelled each stream AVTPDU so that other researchers can design sophisticated experiments.

As part of our previous report, we released a CAN intrusion dataset \cite{Lee2017}. Interestingly, we confirmed that many studies had been conducted using the published dataset. We hope that our intrusion dataset will trigger various studies on automotive Ethernet. Readers can refer to \cite{dataset} to access our automotive Ethernet intrusion datasets.

\subsection{Limitations}
\label{subsec:limitation}
The proposed IDS must receive all stream AVTPDUs for intrusion detection continuously. As our IDS is designed to detect intrusions through continuous stream changes, intermittent traffic reception will cause numerous false positives. To solve this problem, we expect the implemented IDS to act as an AVB listener (see the AVB listener connected to the IDS in \autoref{fig:topology}) because stream AVTPDUs are delivered to all participating AVB listeners.

Through the experiment, we confirmed that our model shows outstanding classification performance. However, the pretrained model exhibits a slight performance degradation when the packet capture environment is changed (from \dataset1 to \dataset2). This is a common problem observed in deep learning models. Future studies can address this in two ways. First, a deep learning model should be designed to be more robust than the proposed model. A common method for achieving this is deploying additional layers in a model; however, this entails a trade-off relationship with the classification speed. Second, a transfer learning technique can be considered to improve classification performance in a new environment. In this way, the pretrained intrusion detection model can be further trained using new inputs (but not trained from scratch). 

The proposed intrusion detection method needs additional hardware for real-time inference. The experiment result implies that our intrusion detection method requires GPU acceleration or a high-end CPU to detect intrusions in real-time. Real-time detection was not possible on small embedded computers such as Raspberry Pi 3, although our 2D-CNN model consisted of relatively few layers. Fortunately, real-time intrusion detection can be performed using a low-power embedded GPU.

\subsection{Remediation strategies}
In this study, intrusion detection was performed under the assumption that an attacker has access to an automotive Ethernet-based IVN. It is important to detect intrusions to protect IVNs. However, it is also essential to make it difficult for an attack to occur and to enable a quick response to an identified attack. In brief, we suggest the following remediation strategies:

\begin{enumerate}
    \item Efforts are needed to reduce the attack surface. A firewall should be installed on known attack surfaces to prevent arbitrary packet injections. In addition, security checks should be performed on nodes that may be subject to external attacks (such as infotainment devices). 
    \item Encryption and authentication protocols should be employed so that passive attackers do not easily understand the internal communication process. For example, MACSec ensures the confidentiality and integrity of communication over automotive Ethernet \cite{Carnevale2018MACsec-BasedBackbones}.
    \item It is possible to prevent bandwidth occupancy (i.e., DoS attacks) by using the stream reservation protocol. This protocol is designed to ensure quality-of-service by reserving and increasing the bandwidth for time-sensitive A/V streams on automotive Ethernet. The reserved bandwidth can prevent communication failures due to high traffic.
    \item Finally, it is necessary to develop intrusion response systems to block an attack as soon as it is identified.
\end{enumerate}

%% file: paper/07_related_work.tex
\section{Related work}
\label{sec:related_work}
Corbett \etal \cite{Corbett2017ASystems} presented a pioneering study for Ethernet-based vehicle networks by analyzing Ethernet protocols designed for vehicle diagnostics, communication patterns, state-of-the-art tools, and attack scenarios. They also proposed a testing framework for an automotive IDS involving parameters, metrics, and automotive-specific challenges. However, their work only provides concepts regarding framework architecture and does not provide results of realistic experiments with an implemented IDS. Consequently, their testing tool needs to be evaluated.

Corbett \etal \cite{Corbett2016AutomotiveChallenge} also conducted an early study that discusses security mechanisms for automotive Ethernet-based IVNs. The authors identified possible manipulation and misuse attacks on automotive Ethernet-based networks, followed by an analysis of network topology and participating devices. The protocols and operating systems designed for vehicles were also introduced.

Readers interested in the current security trends, threats, and intrusion detection methods of IVNs are encouraged to refer to state-of-the-art reviews \cite{Loukas2019, Wu2019} and the references therein. The reviews also describe the constraints, challenges, and characteristics of IDSs for IVNs. This information will be of great help in further research on vehicle intrusion detection methods. However, these reviews lack citations of Ethernet- and automotive Ethernet-specific studies.

Some research was conducted on network packet classification using deep learning techniques. Kim and Anpalagan \cite{Kim2018} adopted a 1D-CNN model to classify encrypted Tor traffic using only the TCP/IP header. Their data processor extracts the first 54 bytes of each payload and then split bytes into nibbles. The nibbles are used as the inputs of the 1D-CNN model. The experimental results show that this model not only performs binary classification well but can also infer specific application types. Wang \etal \cite{Wang2017} converted packets to $28\times28$ sized grey images to classify malware traffic using a CNN model. This CNN model has the advantages of early-stage detection, low false-alarm rate, protocol-independency, automatic extraction of features, and raw traffic data input. On a CAN-based IVN system, Song \etal \cite{Song2019} adopted a deep CNN model--Inception-ResNet--to classify four types of injection attacks within the CAN bus. Their frame builder is designed to represent a sequence of CAN IDs at a bit level. Taylor \etal \cite{Taylor2016AnomalyNetworks} also conducted a study to detect CAN bus attacks using long short-term memory (LSTM) networks.

Changing the nature of the packets and environment degrades performance when a pretrained deep learning model is used for detection. For instance, this degradation is apparent from \autoref{table:result_test}. Moreover, new types of attacks are not properly detected by existing models. Tariq \etal \cite{Tariq2020CANTransferNetwork} applied transfer learning to identify new types of intrusion attacks. They trained a convolutional LSTM network using a DoS attack dataset captured from the CAN bus. Moreover, they tried one-shot learning using only a few fuzzing and replay attacks.

%% file: paper/08_conclusion.tex
\section{Conclusion}
\label{sec:conclusion}
In this paper, we proposed an intrusion detection method that detects AVTP stream injection attacks in automotive Ethernet. To the best of our knowledge, this is the first intrusion detection method for protecting automotive Ethernet-based IVNs. Our approach includes feature generation and a CNN-based intrusion detection model. We design a feature generator that measures state changes of the AVTP stream at the nibble-level within a specific time window. The CNN model extracts features from the input and infers whether a stream AVTPDU is benign or injected. The experimental results show that our model achieved a high detection performance and remarkably high recall for real stream AVTPDUs captured from our BroadR-Reach-based testbed. We prove that our model is suitable for real-time detection in an actual CAV by measuring the time required for inference. Although we limited our experiments to stream AVTPDUs, we will also consider other AVB-related protocols and automotive diagnostics communication protocols in future work. Finally, we have released the dataset collected for this study. We look forward to further studies on automotive Ethernet security based on our AVTP stream dataset.

%% file: main.bbl
\begin{thebibliography}{10}
\expandafter\ifx\csname url\endcsname\relax
  \def\url#1{\texttt{#1}}\fi
\expandafter\ifx\csname urlprefix\endcsname\relax\def\urlprefix{URL }\fi
\expandafter\ifx\csname href\endcsname\relax
  \def\href#1#2{#2} \def\path#1{#1}\fi

\bibitem{Hank2013AutomotiveMobility}
P.~Hank, S.~Muller, O.~Vermesan, J.~Van Den~Keybus,
  \href{http://ieeexplore.ieee.org/xpl/articleDetails.jsp?arnumber=6513795}{{Automotive
  Ethernet: In-vehicle Networking and Smart Mobility}}, in: Design, Automation
  {\&} Test in Europe Conference {\&} Exhibition (DATE), 2013, IEEE Conference
  Publications, New Jersey, 2013, pp. 1735--1739.
\newblock \href {https://doi.org/10.7873/DATE.2013.349}
  {\path{doi:10.7873/DATE.2013.349}}.
\newline\urlprefix\url{http://ieeexplore.ieee.org/xpl/articleDetails.jsp?arnumber=6513795}

\bibitem{ieee1722-2016}
{IEEE P1722 working group},
  \href{https://doi.org/10.1109/IEEESTD.2016.7782716}{{IEEE} standard for a
  transport protocol for time-sensitive applications in bridged local area
  networks} (2016).
\newblock \href {https://doi.org/IEEESTD.2016.7782716}
  {\path{doi:IEEESTD.2016.7782716}}.
\newline\urlprefix\url{https://doi.org/10.1109/IEEESTD.2016.7782716}

\bibitem{Loukas2019}
G.~Loukas, E.~Karapistoli, E.~Panaousis, P.~Sarigiannidis, A.~Bezemskij,
  T.~Vuong, \href{https://doi.org/10.1016/j.adhoc.2018.10.002}{{A taxonomy and
  survey of cyber-physical intrusion detection approaches for vehicles}}, Ad
  Hoc Networks 84 (2019) 124--147.
\newblock \href {https://doi.org/10.1016/j.adhoc.2018.10.002}
  {\path{doi:10.1016/j.adhoc.2018.10.002}}.
\newline\urlprefix\url{https://doi.org/10.1016/j.adhoc.2018.10.002}

\bibitem{Panarotto2018}
F.~Panarotto, A.~Cortesi, P.~Ferrara, A.~K. Mandal, F.~Spoto,
  \href{http://link.springer.com/10.1007/978-3-030-05755-8_12}{{Static Analysis
  of Android Apps Interaction with Automotive CAN}}, in: nternational
  Conference on Smart Computing and Communication (SmartCom), 2018, pp.
  114--123.
\newblock \href {https://doi.org/10.1007/978-3-030-05755-8{\_}12}
  {\path{doi:10.1007/978-3-030-05755-8{\_}12}}.
\newline\urlprefix\url{http://link.springer.com/10.1007/978-3-030-05755-8_12}

\bibitem{Wu2019}
W.~Wu, R.~Li, G.~Xie, J.~An, Y.~Bai, J.~Zhou, K.~Li,
  \href{https://ieeexplore.ieee.org/document/8688625/}{{A Survey of Intrusion
  Detection for In-Vehicle Networks}}, IEEE Transactions on Intelligent
  Transportation Systems 21~(3) (2020) 919--933.
\newblock \href {https://doi.org/10.1109/TITS.2019.2908074}
  {\path{doi:10.1109/TITS.2019.2908074}}.
\newline\urlprefix\url{https://ieeexplore.ieee.org/document/8688625/}

\bibitem{dataset}
S.~Jeong, B.~Jeon, B.~Jung, H.~K. Kim, Automotive ethernet intrusion datasets,
  \url{https://github.com/seonghoony/autoeth-intrusion-dataset} (2020).

\bibitem{autoethdefinitiveguide}
B.~Metcalfe, C.~M. Kozierok, C.~Correa, R.~B. Boatright, J.~Quesnelle,
  M.~Holden, K.~Irving, Automotive Ethernet, Intrepid Control Systems, 2014.

\bibitem{Alabady2008DesignServer}
S.~A.~J. Alabady, \href{http://ieeexplore.ieee.org/document/4530276/}{{Design
  and Implementation of a Network Security Model using Static VLAN and AAA
  Server}}, in: 2008 3rd International Conference on Information and
  Communication Technologies: From Theory to Applications, IEEE, 2008, pp.
  1--6.
\newblock \href {https://doi.org/10.1109/ICTTA.2008.4530276}
  {\path{doi:10.1109/ICTTA.2008.4530276}}.
\newline\urlprefix\url{http://ieeexplore.ieee.org/document/4530276/}

\bibitem{vyncke2006layer2sec}
E.~Vyncke, Layer 2 security,
  \url{https://www.cisco.com/c/dam/global/da_dk/assets/docs/security2006/Security2006_Eric_Vyncke_2.pdf}
  (2006).

\bibitem{rezvani2018hacking}
D.~Rezvani, Hacking automotive ethernet cameras,
  \url{https://argus-sec.com/hacking-automotive-ethernet-cameras/} (2018).

\bibitem{ioffe15batch}
S.~Ioffe, C.~Szegedy,
  \href{http://jmlr.org/proceedings/papers/v37/ioffe15.pdf}{Batch
  normalization: Accelerating deep network training by reducing internal
  covariate shift}, in: International Conference on International Conference on
  Machine Learning, 2015, pp. 448--456.
\newline\urlprefix\url{http://jmlr.org/proceedings/papers/v37/ioffe15.pdf}

\bibitem{JMLR:v15:srivastava14a}
N.~Srivastava, G.~Hinton, A.~Krizhevsky, I.~Sutskever, R.~Salakhutdinov,
  \href{http://jmlr.org/papers/v15/srivastava14a.html}{Dropout: A simple way to
  prevent neural networks from overfitting}, Journal of Machine Learning
  Research 15 (2014) 1929--1958.
\newline\urlprefix\url{http://jmlr.org/papers/v15/srivastava14a.html}

\bibitem{Kingma2014}
D.~P. Kingma, J.~Ba, \href{http://arxiv.org/abs/1412.6980}{{Adam: A Method for
  Stochastic Optimization}}, arXiv Preprint (12 2014).
\newline\urlprefix\url{http://arxiv.org/abs/1412.6980}

\bibitem{chollet2015keras}
F.~Chollet, et~al., Keras, \url{https://keras.io} (2015).

\bibitem{Lee2017}
H.~Lee, S.~H. Jeong, H.~K. Kim,
  \href{https://ieeexplore.ieee.org/document/8476919/}{{OTIDS: A Novel
  Intrusion Detection System for In-vehicle Network by Using Remote Frame}},
  in: 2017 15th Annual Conference on Privacy, Security and Trust (PST), Vol.~5,
  IEEE, 2017, pp. 57--5709.
\newblock \href {https://doi.org/10.1109/PST.2017.00017}
  {\path{doi:10.1109/PST.2017.00017}}.
\newline\urlprefix\url{https://ieeexplore.ieee.org/document/8476919/}

\bibitem{Carnevale2018MACsec-BasedBackbones}
B.~Carnevale, L.~Fanucci, S.~Bisase, H.~Hunjan, {MACsec-Based Security for
  Automotive Ethernet Backbones}, Journal of Circuits, Systems and Computers
  27~(5) (2018) 1--17.
\newblock \href {https://doi.org/10.1142/S0218126618500822}
  {\path{doi:10.1142/S0218126618500822}}.

\bibitem{Corbett2017ASystems}
C.~Corbett, T.~Basic, T.~Lukaseder, F.~Kargl, {A testing framework architecture
  concept for automotive intrusion detection systems}, Lecture Notes in
  Informatics (LNI), Proceedings - Series of the Gesellschaft fur Informatik
  (GI) P-269 (2017) 89--102.

\bibitem{Corbett2016AutomotiveChallenge}
C.~Corbett, E.~Schoch, F.~Kargl, P.~Felix,
  \href{http://www.gi.de/service/publikationen/lni/}{{Automotive ethernet:
  Security opportunity or challenge?}}, in: Lecture Notes in Informatics (LIN),
  2016, pp. 45--54.
\newline\urlprefix\url{http://www.gi.de/service/publikationen/lni/}

\bibitem{Kim2018}
M.~Kim, A.~Anpalagan, \href{https://ieeexplore.ieee.org/document/8569113/}{{Tor
  Traffic Classification from Raw Packet Header using Convolutional Neural
  Network}}, in: 2018 1st IEEE International Conference on Knowledge Innovation
  and Invention (ICKII), no. January, IEEE, 2018, pp. 187--190.
\newblock \href {https://doi.org/10.1109/ICKII.2018.8569113}
  {\path{doi:10.1109/ICKII.2018.8569113}}.
\newline\urlprefix\url{https://ieeexplore.ieee.org/document/8569113/}

\bibitem{Wang2017}
{Wei Wang}, {Ming Zhu}, {Xuewen Zeng}, {Xiaozhou Ye}, {Yiqiang Sheng},
  \href{http://ieeexplore.ieee.org/document/7899588/}{{Malware traffic
  classification using convolutional neural network for representation
  learning}}, in: 2017 International Conference on Information Networking
  (ICOIN), IEEE, 2017, pp. 712--717.
\newblock \href {https://doi.org/10.1109/ICOIN.2017.7899588}
  {\path{doi:10.1109/ICOIN.2017.7899588}}.
\newline\urlprefix\url{http://ieeexplore.ieee.org/document/7899588/}

\bibitem{Song2019}
H.~M. Song, J.~Woo, H.~K. Kim,
  \href{https://doi.org/10.1016/j.vehcom.2019.100198}{{In-vehicle network
  intrusion detection using deep convolutional neural network}}, Vehicular
  Communications 21 (2020) 100198.
\newblock \href {https://doi.org/10.1016/j.vehcom.2019.100198}
  {\path{doi:10.1016/j.vehcom.2019.100198}}.
\newline\urlprefix\url{https://doi.org/10.1016/j.vehcom.2019.100198}

\bibitem{Taylor2016AnomalyNetworks}
A.~Taylor, S.~Leblanc, N.~Japkowicz,
  \href{http://ieeexplore.ieee.org/document/7796898/}{{Anomaly Detection in
  Automobile Control Network Data with Long Short-Term Memory Networks}}, in:
  2016 IEEE International Conference on Data Science and Advanced Analytics
  (DSAA), IEEE, 2016, pp. 130--139.
\newblock \href {https://doi.org/10.1109/DSAA.2016.20}
  {\path{doi:10.1109/DSAA.2016.20}}.
\newline\urlprefix\url{http://ieeexplore.ieee.org/document/7796898/}

\bibitem{Tariq2020CANTransferNetwork}
S.~Tariq, S.~Lee, S.~S. Woo,
  \href{https://dl.acm.org/doi/10.1145/3341105.3373868}{{CANTransfer - Transfer
  Learning based Intrusion Detection on a Controller Area Network using
  Convolutional LSTM Network}}, in: Proceedings of the 35th Annual ACM
  Symposium on Applied Computing, ACM, New York, NY, USA, 2020, pp. 1048--1055.
\newblock \href {https://doi.org/10.1145/3341105.3373868}
  {\path{doi:10.1145/3341105.3373868}}.
\newline\urlprefix\url{https://dl.acm.org/doi/10.1145/3341105.3373868}

\end{thebibliography}
